# Exposing Hate:
# Understanding Anti-Immigration Sentiment Spreading on Twitter

**Andrea Nasuto\*, Francisco Rowe**

Geographic Data Science Lab, University of Liverpool, UK

**Abstract:**

Immigration is one of the most salient topics in public debate. Social media heavily influences opinions on immigration, often sparking polarized debates and offline tensions. Studying 220,870 immigration-related tweets in the UK, we assessed the extent of polarization, key content creators and disseminators, and the speed of content dissemination. We identify a high degree of online polarization between pro and anti-immigration communities. We found that the anti-migration community is small but denser and more active than the pro-immigration community with the top 1% of users responsible for over 23% of anti-immigration tweets and 21% of retweets. We also discovered that anti-immigration content spreads also 1.66 times faster than pro-immigration messages and bots have minimal impact on content dissemination. Our findings suggest that identifying and tracking highly active users could curb anti-immigration sentiment, potentially easing social polarization and shaping broader societal attitudes toward migration.

**Keywords**: Immigration sentiment, Social media, Polarization, Online Hate, Online Content Speed, Bots

# 1. Introduction

Public sentiment around immigration is a key societal issue. It polarizes and defines political agendas, and policies and influences social manifestations (Alesina & Tabellini, 2022). Extreme social divisions on immigration harm minorities and diminish social cohesion creating conditions for violence (Wahlström & Törnberg, 2019), reduced labor force participation (Becker, 2010), and less sense of belonging among migrants (Tyrberg, 2023). These frictions are exemplified by political outcomes with clear anti-immigration agendas like Brexit and the ascent of far-right parties (Dennison & Geddes, 2019) as well as the formation of pro-immigration movements like 'Refugees Welcome' (Schiffauer, 2019).

Online social media networks have become a prominent space to express public opinions about migration and migrants (Ekman, 2019). They represent a public forum where people can express their opinions openly and these can spread rapidly and globally (Brown et al., 2007). Unlike traditional methods, such as community meetings or traditional media, social media provides an easy, quick, and remotely accessible open forum where people can freely share their views on immigration. The speed and global reach of these platforms allow opinions to spread rapidly transcending geographical limitations. However, challenges arise from this dissemination of immigration-related content, including the spread of misinformation and the amplification of polarized views. Nevertheless, social media offers an accessible and dynamic medium for engaging in dialogue, contributing to public discourse, and raising awareness of migration-related challenges.

The existing knowledge about the dissemination of immigration sentiment through social media is currently limited. Previous research has mainly focused on how alternative digital sources, including social media data, can be utilized to track immigration sentiment (Freire-Vidal et al., 2021). However, these studies have often been confined to specific events (Arcila-Calderón et al., 2021) or exploring monitoring techniques (Rowe et al., 2023). It has been observed that social media platforms play a role in promoting anti-immigration sentiment by efficiently spreading misinformation and polarizing content related to migrants (Bognár & Szakács, 2021). Such content can influence public opinion by amplifying fears and biases against migrants (Butcher & Neidhardt, 2020). Additionally, social media's significance as a platform for news consumption contributes to shaping public sentiment on immigration, as it has been shown to reinforce existing biases and polarization (Del Vicario et al., 2017). Some studies have examined the extent of polarization in online debates about immigration within particular countries (Bursztyn et al., 2019; Vilella et al., 2020), as well as the involvement of bots in disseminating anti-immigration content (Bastos & Mercea, 2019). However, there are still important dimensions of the online migration network that have not been extensively explored.

To develop a better understanding of online immigration sentiment, we need to investigate three key dimensions: the extent of the polarization, the key sources and the speed of the immigration-related content.

Other research fields have more broadly examined these dimensions, such as investigating the speed of online misinformation (Vosoughi et al., 2018), analyzing the polarization of political discourse in online platforms (Kubin & von Sikorski, 2021), and identifying the primary sources of content during elections (Grinberg et al., 2019).

Leveraging natural language processing (NLP) methods and social network science (SNA), we aim to:
- Determine the extent of polarization of social media immigration sentiment;
- Identify the key producers and spreaders of social media immigration-related content;
- Measure the speed at which this content is disseminated through social media immigration communities.

Based on 220,870 tweets collected by Rowe et al. (2021) for the UK, the research examines the immigration sentiment between December 1st 2019 to April 30th 2020 which includes the UK general election (December 13th 2019) and the first wave of the COVID-19 pandemic in the UK. Twitter is an alternative source of data to measure social phenomena including immigration sentiment (Freire-Vidal et al., 2021), while its usage presents its own challenges including multiple biases in its userbase and data quality (Flores, 2017). The granularity and the scale of the data make Twitter data an excellent source to study almost in real-time the public sentiment on immigration and potentially design effective policies to halt online abuses and hate.

The rest of the paper is structured in five sections. The next section presents existing work on immigration sentiment on online social networks and patterns of the general attitudes towards immigration. It also describes the role of social media in shaping public opinions analyzing social media polarization, users' heterogeneity in producing and spreading content and the speed at which online content circulates. To this end, we draw on research on the digital online spread and discussions of other research areas including misinformation and political polarization. We then introduce the data and methods used in Section Two before presenting the results in Section Three. Next, we discuss our findings in Section Four, and we conclude Section Five by delineating the policy implications and avenues for future research of our research.

## 2. Background

Migration sentiment is a crucial social issue that influences integration, discrimination, and human rights. Positive sentiment fosters social cohesion, while negative sentiment hinders integration and perpetuates discrimination. Understanding public opinion on immigration is essential for inclusive policies including leveraging the economic potential of migration (Callens, 2015; Qi et al., 2021). Immigration has emerged as a significant challenge in various countries, often depicted as a major concern that influences political narratives, as exemplified by events like Brexit and the ascent of far-right parties (Dennison & Geddes, 2019). Extreme right-wing narratives have often depicted migrants as the 'scapegoats' for all the challenges within a country creating the conditions for a 'moral panic' towards migrants and immigration (Walsh & Hill, 2023).

Immigration has also been a prominent and enduring concern for the UK population over the past two decades, consistently ranking among the top three most important issues for British voters (Ipsos, 2023b). This issue has been linked to a concerning rise in racially motivated hate crimes (Allen & Zayed, 2022). Survey polls also indicate a growing negative opinion on immigration while the significance of immigration has decreased since the EU 2016 referendum in the UK (Ipsos, 2023a).

Assessing immigration has remained a complex endeavor, aimed at offering a real-time understanding of the multifaceted issues associated with immigration. Established research methods of data collection have been mostly based on time-consuming, time-sparse, expensive, and spatially coarse surveys. These methods also rely on pre-defined questions which offer insights into specific issues of the immigration process. Empirical research has mostly relied on a question asking respondents for their position in relation to raising or decreasing immigration quotas (IOM, 2015). The view of immigration based on these questions is less positive than if we consider other dimensions of the impact of immigration on culture and diversity (Ueffing et al., 2015). Alternative approaches can overcome some of the key drawbacks of well-established data sources. Social media has emerged as a unique source to track public sentiment on immigration (Bosco et al., 2022). Social media platforms capture content on immigration but also can serve a role in shaping public opinions around immigration (Gillespie, 2020; Walsh & Hill, 2023).

Social media has become a de facto public forum for exchanging opinions worldwide, including immigration. Social media can influence the public debate on immigration through a series of mechanisms and features of the platforms.

Firstly, social media platforms serve as a primary source of news and information for many individuals (Newman et al., 2023). Studies have shown that unfavorable news coverage about migration can reduce the acceptance of asylum seekers by imposing more restrictive policies (Koch et al., 2020). This coverage can be enhanced by social media further reinforcing this news consumption pattern (Del Vicario et al., 2017). Indeed, the information presented on social media platforms can be fragmented and biased, leading to potential polarization and the reinforcement of existing attitudes toward migrants (Vilella et al., 2020).

Secondly, social media enables the rapid spread of content, especially emotionally charged narratives. This rapid dissemination of online emotional responses can quickly trigger violent actions on migrants (Bognár & Szakács, 2021). Indeed, social media platforms have facilitated the rise of online activism and mobilization around migration issues. Campaigns, hashtags, and user-generated content on social media can raise awareness, shape public discourse, and influence public opinion (Acemoglu et al., 2017; Enikolopov et al., 2020). Activist groups and organizations leverage social media to amplify their messages and mobilize support for their causes. For example, both pro-immigration movements, such as 'Refugees Welcome' and anti-immigration movements such as the 'Leave' Brexit have been facilitated by social media engagement (Dennison & Geddes, 2019; Schiffauer, 2019).

Social media platforms are also known for spreading misinformation and disinformation related to migration (Bognár & Szakács, 2021). False or misleading narratives on migrants can shape public opinion by fueling fears, stereotypes, and biases (Butcher & Neidhardt, 2020). Research has shown that online misinformation can have a significant impact on public perceptions and attitudes towards immigration (Bosco et al., 2022).

Although there is an increasing number of empirical studies analyzing public opinion on migration, the research in this area remains limited. Existing research mostly focuses on exploring the potential of the use of social media to analyze public opinion on migration (Bosco et al., 2022) or assessing the prevalence of positive and negative sentiment towards immigration (Gualda & Rebollo, 2016; Rowe et al., 2023; Sanguinetti et al., 2018). Temporally, a longer time span analysis is missing since most of the previous research has explored how immigration sentiment takes shape during notable events in Spain (Arcila-Calderón et al., 2021) and in the UK (Williams & Burnap, 2015). As a result, an enhanced understanding of the key structural elements of the online debate on immigration is required.

This includes identifying the extent of the network polarization, the size of a network, the key actors involved in the debate and how content disseminates. By analyzing these elements, we can gain insights into the dynamics of information flow, social interactions, and the formation of sentiment.

The polarization in the online debate on immigration has received limited research attention despite its societal significance. Polarization in the online immigration sentiment has been analyzed in Italy (Vilella et al., 2020) and Russia (Bursztyn et al., 2019). Vilella et al. (2020) showed that the Italian immigration debate on Twitter is highly polarized with low level of interactions between communities on the opposite side of the spectrum. On the other hand, Ziems et al. (2020) have shown how users involved in anti-Asian hate and counterhate online speech during COVID-19 in the US tend to have interconnected interactions without confining themselves to isolated groups. The online public discourse on immigration can be significantly influenced by a few influential users. Yet, we have limited knowledge on how key users could shape the online debate on immigration. The speed at which different immigration-related content spreads within a network is also relevant. Empirical evidence shows that the quick spread of online misinformation about migrants can rapidly lead to physical violence (Bognár & Szakács, 2021). Yet no study has investigated the pace at which immigration content spreads online and assessed the differences between anti and pro-immigration content.

Bots could play an active role in shaping the online debate on immigration by both acting as primary sources in a network of users and accelerating the pace of content dissemination. Indeed, online social networks have also witnessed the emergence of non-human users otherwise known as 'bots' which can increase polarization and enhance the spread of misinformation (Bessi & Ferrara, 2016). Bastos and Mercea (2019) have studied the role of bots in spreading pro-Brexit campaign messages highlighting how bots can be used to leverage quasi-fake news mostly around immigration. However, the specific impact of bots on the spread of immigration-related content has yet to be tested.

While the existing literature on immigration lacks the study of these key dimensions on social media, previous studies have explored these characteristics across multiple research areas including the online debate on politics (Kubin & von Sikorski, 2021), COVID-19 (Jiang et al., 2020), climate change (Falkenberg et al., 2022) as well as misinformation (Vosoughi et al., 2018) (Vosoughi et al., 2018), fake news (Grinberg et al., 2019) and conspiracy theories (Del Vicario et al., 2016).

Social media have been publicly blamed for increasing societal polarization by fueling divisions through the creation of 'echo chambers' or 'filter bubbles' which reinforce existing ideas and biases by carefully excluding different voices from a certain debate (Sunstein, 2017). On the other hand, other research suggests quite the opposite. Social media platforms expose users to a multiplicity of perspectives, including those from the opposite spectrum, thereby instigating negative and hateful interactions (Bail et al., 2018; Törnberg, 2022). A practice also known as 'ratioing' or 'boo and cheer' (Bartlett & Norrie, 2015) which can be centered around key polarizing users within a debate.

Indeed, a small number of key users can have an outsized impact in shaping an online debate. Previous studies have shown that the so-called 'influencers' have a substantial role in fostering extremism in the user base rather than the content per se (Becker et al., 2019). Similarly, (Grinberg et al., 2019) discovered that 0.1% of the users shared 80% of the fake news on Twitter during the 2016 US presidential election. Twitter itself has used this approach to highlight the disproportionate contribution of users across the platform (Reuters, 2022). The pace at which online content on immigration spreads could also vastly change the patterns in the debate as well as trigger uncontrolled offline reactions (Bognár & Szakács, 2021). Research on misinformation has shown falsehood content spreads faster (Vosoughi et al., 2018) while conspiracy-based content seems to spread slower than science-based information (Del Vicario et al., 2016). Evidence suggests that hateful content has higher speed and higher spread on social media platforms (Mathew et al., 2019).

The impact of the bots on the spread of online content. Regarding the relationship between the speed of Twitter content and non-human users, Shao et al. (2018) discovered a positive correlation, although the significance of this finding has been challenged by other researchers (Vosoughi et al., 2018). Nonetheless, bots appear to have an amplifying role in disseminating divisive and hateful content (Stella et al., 2019;

Uyheng et al., 2022) and reshaping the dynamics of the debate (Bovet & Makse, 2019; Caldarelli et al., 2020).

## 3. Materials and methods

### 3.1 Data

The research leverages Twitter data to analyze online public sentiment toward immigration. The data were collected by Rowe et al. (2021) across five countries (Germany, Italy, Spain, the UK, USA) between December 1st, 2019 and April 30th, 2020 for a total of 30.39 million data points. This study uses only the UK data, resulting in a total of 220,870 tweets. Rowe et al. (2021) used a methodology to collect a curated sample of tweets leveraging the Premium Twitter API. Each day, 500 tweets across three different query types (hashtag, account and key terms) were obtained for a total of 1500 tweets per day. Data were processed by Rowe et al. (2021): (1) removing duplicated retweets, and non-relevant migration-related tweets (e.g. concerning bird migrations); (2) converting emojis and hashtags into the text; and, (3) removing account usernames, URLs, and hyperlinks.

### 3.2 Methods

Our analysis involves five stages. Firstly, we classified each tweet based on their standing towards immigration using a fine-tuned BERT transformer. Secondly, we label the users as being pro- or anti-immigration based on the proportion of anti- or pro-immigration tweets shared. Thirdly, we use social network science (SNA) methods to understand the structure of the users involved in the Twitter debate on migration. Specifically, we measure the strength of the polarization in the debate, the density of the pro- and anti-immigration users' networks and quantify the strength of the relationship between spreaders as well as producers of anti- and pro-immigration content. Fourth, we identified who are the key spreaders and producers of anti and pro-immigration content by calculating the top 1% by the number of retweets to identify the spreaders and the top 1% by the number of tweets generated to identify the producers. Fifth, we wanted to understand the differences in speed across types of content. Thus, we calculated two complementary cumulative distribution functions (CCDFs) to assess the difference in the number of retweets between anti- and pro-immigration content. We also calculated the median time for an anti- and pro-immigration tweet to reach a certain number of retweets. Sixth, we identified bots i.e. non-human users within the network to understand what is their impact on content dissemination. Next, each stage is described in detail.

#### 3.2.1 Text classification

We built a text classifier to identify the different standings toward immigration in our Twitter dataset. A text classifier uses artificial intelligence and machine learning to automatically identify different types of content processing large amounts of data with speed and accuracy (Minaee et al., 2021). The research uses a pre-trained Bidirectional Encoder Representations from Transformers (BERT) to build a custom-made text classifier that has been fine-tuned on a random subset of manually labeled data.

BERT is a deep learning architecture developed by researchers at Google (Devlin et al., 2018) that is pre-trained on a large unlabeled text corpus. BERT constitutes a new class of methods in natural language processing which have also been known as transformers. Transformers constitute the current 'state of the art' in the NLP methods (Kovaleva et al., 2019) which are the core architecture of large language models. Transformers have been previously used to detect hate speech, outperforming previous methods (Mozafari et al., 2019; Pota et al., 2021; Sun et al., 2019). Indeed, transformers can be tuned for specific tasks i.e. immigration sentiment detection (Ziems et al., 2020).

To fine-tune the transformer, we randomly sampled 1000 tweets from our dataset. We manually labeled them using four exclusive categories: 'pro-immigration', 'anti-immigration', 'neutral', and 'unclassified': 'pro-immigration' labels are used to describe the content of tweets when they express positive opinions towards immigration; 'anti-immigration' labels are used to describe tweets expressing negative opinions towards immigration; 'neutral' labels are used to identify content with no clear sentiment towards immigration; and, 'unclassified' labels are used to categorize content which cannot be classified into one of the categories above, either because the tweets are not completely related to the immigration debate (e.g. "Question: Has anyone successfully attempted a Windows vCenter 6.5 (with embedded PSC) to vCSA 6.5 migration? /cc @kev_johnson") or cannot be assessed in the context (e.g. "@RSPCA_official @ukhomeoffice Maybe @EventbriteGB could assist?").

The training dataset is unbalanced across the four categories thus we oversampled tweets labeled 'neutral' and 'anti-immigration' and undersampled tweets labeled as 'unclassified' in the final training dataset. This dataset is then used to fine-tune our classifier. The optimal learning rate is iteratively determined to optimize accuracy and minimize the loss and it is visually selected by analyzing the auto-generated loss plot through a built-in ktrain library function. The epochs, batch size, the max level of tokenization and the max number of features are all selected through an iterative process building multiple classifiers and progressively selecting the best performers by looking at the F1 scores across variables and the general AUC-ROC score.

### 3.2.2 User classification

We classified the users based on their tweets shared. A user is labeled as being pro-immigration if more than 50% of her tweets are pro-immigration, while the opposite holds true for anti-immigration. A boolean numeric variable is assigned to all tweets labeled either anti- (0) or pro-immigration (1). Each user is thus identified as being pro-immigration or anti-immigration by calculating the average of the newly assigned boolean variable.

### 3.2.3 Network Analysis

We employed Social Network Analysis (SNA) to examine the structure of the Twitter discourse regarding migration. In this analysis, we designated Twitter users as nodes and represented their interactions through retweets as edges. Nodes in this context serve as both producers and spreaders of content, with edges indicating the flow of content from one node to another, establishing a directed network. Each node has a degree that quantifies the number of edges associated with that node. A higher node degree signifies a higher number of retweets. Within our directed network, we further distinguish nodes based on their in-degree and out-degree values. In our study, a node with a high in-degree signifies a user that frequently retweets, essentially a prominent content spreader. Conversely, a node with a high out-degree indicates a user who generates content that is regularly and widely retweeted, essentially a prolific content producer. The resulting directed network comprises 34,063 nodes and 48,883 edges. Additionally, we identified two distinct subnetworks: the pro-immigration network and the anti-immigration network. These subnetworks were discerned by evaluating users' stances on immigration.

We assessed the network's structure, namely the polarization in the network, the density of the anti- and pro-immigration networks and the strength of the interactions between producers as well as spreaders in the anti- and pro-immigration network. We computed four metrics: the attribute assortativity coefficient, in-degree assortativity coefficient, out-degree assortativity coefficient and edge density.

We calculated the attribute assortativity coefficient as:

$$r\text{attribute} = \frac{\sum_i e_{ii} - \sum_i a_i b_i}{1 - \sum_i a_i b_i}$$

where $e_{ii}$ is the probability of an edge (a retweet) between two nodes (users) which have both a given standing towards immigration, $a_i$ is the probability that an edge has as origin a node with given standing towards immigration and $b_i$ is the probability that an edge has as destination a node with value $i$.

We also calculated the in-degree (out-degree) assortativity coefficient as:

$$r\text{in/out} - \text{degree} = \frac{\sum_{i,j} P_i P_j (e_{ij} - a_i b_j)}{\sigma_a \sigma_b}$$

where $P_i P_j (e_{ij} - a_i b_j)$ calculates the product of node properties and their differences for each pair of nodes (users) with specific in-degree values. The summation $\sum_{i,j}$ computes the sum of these products across all pairs of nodes with their respective in-degree properties. $\sigma_a \sigma_b$ in the denominator helps normalize the result, dividing by the product of the standard deviations of the distributions of in-degrees. Symmetrically, the same equation calculates the out-degree assortativity coefficient but only including out-degree edges.

Finally, we calculated the edge density:

$$d = \frac{m}{n(n-1)}$$

where $m$ is the number of edges (retweets) in the users' network and $n$ is the number of nodes (users). The edge density measures the level of interconnectedness within the network based on the retweets.

The assortativity coefficient is a measure to assess homophily in a network. Homophily captures the tendency of individuals to form connections sharing similar attributes, such as opinions or beliefs. In the context of polarization, attribute assortativity can be used to measure the tendency of nodes in a network to connect with others sharing similar opinions or beliefs like standing towards immigration. The more assortative in a network, the more polarized the network is. The attribute assortativity coefficient is calculated as the Pearson correlation coefficient by comparing the observed connections between nodes with a similar standing towards immigration to what would be expected in a random network with the same degree distribution. We also want to understand how spreaders and producers engage within the debate. We calculate the in-degree and out-degree assortativity coefficients. The in-degree assortativity coefficient measures the tendency of nodes to connect with other nodes with a similar number of incoming connections i.e. number of retweets shared. Similarly, the out-degree assortativity coefficient measures the tendency of nodes to connect with users having a similar number of outgoing connections i.e. the number of retweets received. The three assortativity measures were calculated as Pearson correlation coefficients degree between pairs of linked nodes ranging from -1 to 1. An assortative coefficient closer to 1 indicates a perfectly assortative network. On the other hand, a coefficient of -1 indicates disassortative networks. Edge density measures how interconnected are the users within a network. It is calculated by dividing the number of edges present in the network by the total number of potential edges that could exist between its nodes. This metric is often used to assess the sparsity of a network. We considered a network with low edge density to be sparse, indicating that users are not extensively connected through tweets and retweets, but rather they are relatively isolated. On the other hand, a network with high density has more homogenous interactions signaling more engagement within a community.

### 3.2.4 Key sources

We defined key producers and spreaders of Twitter content. A producer is defined as a user generating Twitter content. A spreader is a user who shares (retweets) someone else's content. Users can be both a producer and spreader of content. To define key producers and spreaders, we first identified the number of users generating content (i.e. producers) and the number of users sharing content (i.e. spreaders) and selected the top 1% of users in these two groups by both the count of tweets and retweets. We carried out an analysis of the networks of users with pro and anti-migration stances and compared this to the remaining 99% of users in each group.

### 3.2.5 Content speed and cascade analysis

We built tweet cascades to analyze the speed and reach of migration-related content in our user networks. A tweet cascade is generated using information of a tweet and subsequent retweets thus capturing the history of a tweet and its dissemination on Twitter. Using tweet cascades, we can determine the pace of dissemination for a tweet in the network measuring the time taken to reach a given number of retweets. The size of a tweet cascade can also be measured to capture its reach; that is, the number of times an original tweet was retweeted. To visually represent tweet cascades, we used a complementary cumulative distribution function (CCDF). The CCDF helps to show the fraction of users (either pro or anti-immigration) with a certain number of retweets.

$$CCDF(x) = P(X \geq x)$$

By displaying the CCDF on a chart, we can graphically illustrate the distribution of retweet counts by showing the probability of observing a certain number of retweets or more by type of content i.e. anti- or pro-immigration. The y-axis with CCDF probability is logged given the skewed distribution of the cascades. Bots are subsequently removed from the data and the CCDFs are recalculated to understand their impact.

### 3.2.6 Bot analysis

We also identified non-human actors i.e. bots and how our results are impacted by the content generated by those. Intuitively, we identified and removed accounts as bots to assess how and if our results changed in any statistically significant way. We employed Birdspotter, a Python library created by Ram et al. (2021), to identify bots. Birdspotter processes raw JSON tweet data and generates a 'botness' score, which helps determine the likelihood of an account being a bot. We identified users as bots if they have a botness probability higher than 90%.

# 4. Results

## 4.1 Text classifier results and performance

The accuracy of the text classifier changed across the different labels. F1 scores for 'Unclassified' and 'Neutral' tweets were respectively 0.73 and 0.75 while for the 'Anti-immigration' and 'Pro-immigration' labels 0.55 and 0.64. Table 1 shows the tweets in the dataset as labeled by the text classifier. Most of the tweet content in our sample was classified as 'Unclassified' (36.1%), followed by 'pro-migration' tweets (27.9%) and anti-migration tweets (19.7%). A minority of the tweets are considered neutral (16.1%).

**Table 1.** Tweets (excluding retweets) labeled by the text classifier based on immigration stance

| Label | Tweets | Percent |
| --- | --- | --- |
| Unclassified | 60,417 | 36.14% |
| Pro-migration | 46,737 | 27.96% |
| Anti-migration | 32,980 | 19.73% |
| Neutral | 27,042 | 16.18% |
| Total | 167,176 | 100.00% |

## 4.2 The structure of the users' network

Figure 1 shows our retweet user networks classified as their stances towards migration. Each node has a size proportional to its degree i.e. the number of retweets it has. Figure 1 reveals the existence of two distinct communities with limited interaction between them. The pro-immigration network is 1.69 times larger than the anti-immigration network. Compared to the pro-migration network (in blue), the anti-migration network (in red) is denser and smaller in size. An attribute assortativity coefficient of 0.81 suggests that the level of polarization between these users' networks is high. Users with a similar standing towards migration tend to exclusively retweet content produced by users with similar migration sentiment. Additionally, the anti-immigration network is 2.8x denser than the pro-immigration network by measuring the edge density. A higher network density within the anti-immigration network indicates a strong level of connectivity and engagement among its users.

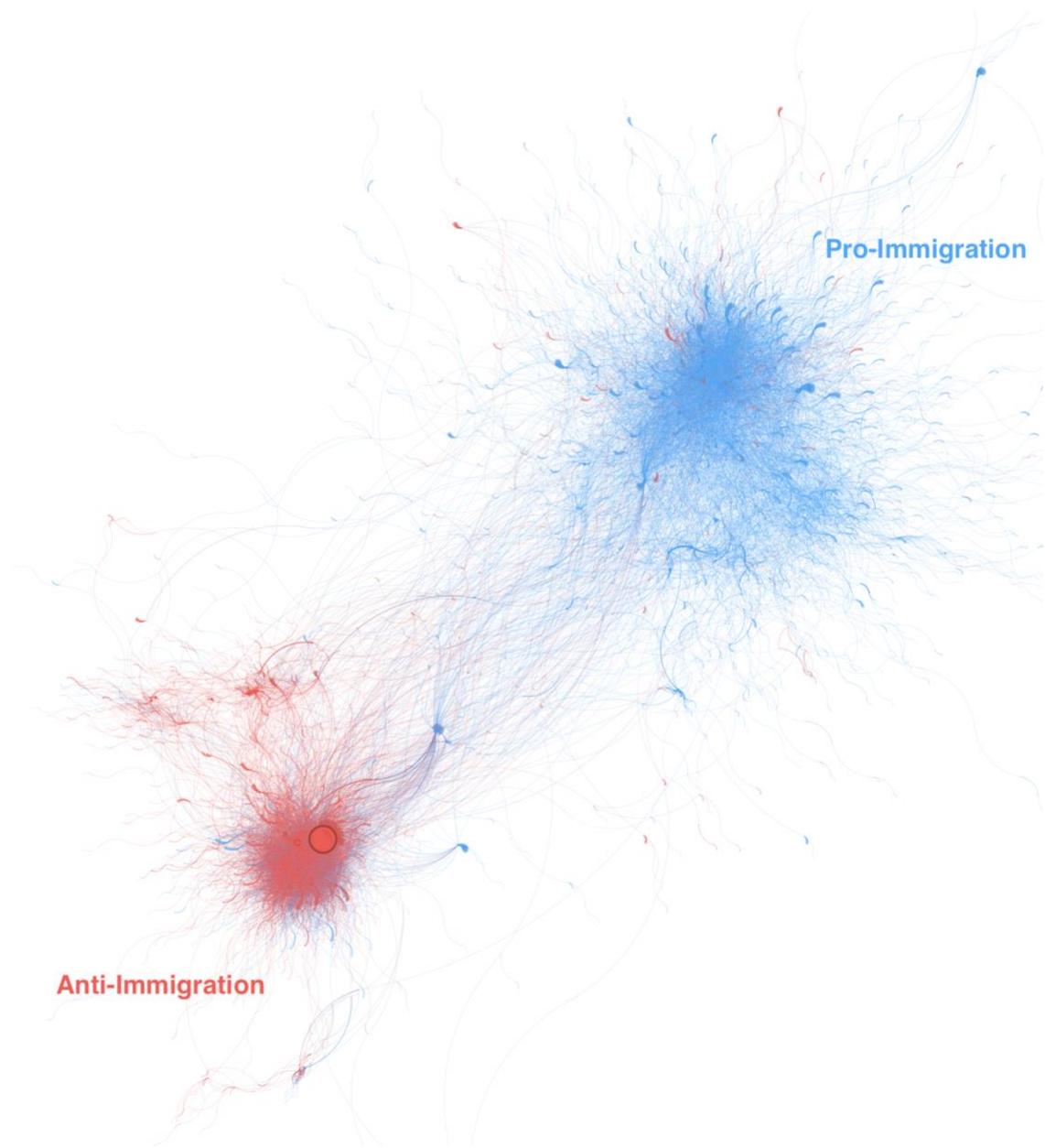

**Figure 1.** Users Network. Retweets directed network of anti-migration (in red) and pro-migration (in blue). Each node is a user and edges are retweets between a source (user creating the original tweet) and a target (user retweeting). The size of the node is proportional to the number of degrees (both in and out) each node has.

The type of engagement also changes across the two communities and the type of users, respectively producers and spreaders. A positive in-degree assortativity coefficient suggests that users who frequently retweet content are inclined to interact with other similarly active retweeters, and vice versa. The anti-immigration community has a higher in-degree coefficient (0.31) compared to the pro-immigration community (0.06). This indicates a distinctive pattern in the anti-immigration community compared to the pro-immigration community. Conversely, both communities exhibit out-degree assortativity coefficients close to 0, indicating a lack of a clear assortative or disassortative network structure in terms of out-degrees.

### 4.3 Key producers of the anti (pro) immigration content

Table 2 shows key metrics to assess the role of the top 1% of users by the number of tweets published (excluding retweets) compared to the bottom 99%. Table 2 also reports the top 1% of producers by their stance on migration. The top 1% of producers account for a disproportionate amount of content shared, particularly among anti-immigration users. The table reveals that key producers across both stances create on average 18.9x more tweets than the remaining 99% of users. The leading producers of anti-immigration content create 23.18% of the total tweets against immigration. In contrast, the top 1% producers of pro-immigration content account for a total of 11.69% of the total tweets supporting immigration.

We also analyzed the presence of bots in the discussion of migration on Twitter. Our results suggest that only a small proportion of top producers are bots (1.5%) across both anti- and pro-immigration users. On the other hand, a higher share (4.12%) of the 99% of producers are identified as bots.

**Table 2** - Producers grouped by quantile and immigration sentiment in the top 1% of the users

|  | Users | Tweets | Tweets (%) | Tweets by User | Botness |
| --- | --- | --- | --- | --- | --- |
| **Top 1%** | 266 | 7,770 | 16.02% | 29.21 | 0.2 |
| **Anti-immigration** | 129 | 4,238 | 23.18% | 32.85% | 0.26 |
| **Pro-immigration** | 137 | 3,532 | 11.69% | 25.78% | 0.15 |
| **Bottom 99%** | 26,396 | 40,724 | 83.98% | 1.54 | 0.4 |

*Note:* The top 1% of producers are selected according to the total number of tweets. Within this top 1%, we identify users classified as 'Anti-immigration' and 'Pro-immigration.' Users' and 'Tweets' columns are counts. 'Tweets by user' is the total count of tweets by category divided by the number of users. 'Botness' is the mean value across the users within that category.

### 4.4 Key spreaders of the anti (pro) immigration content

Table 3 shows key metrics for the top 1% spreaders by the number of retweets compared to the remainder 99%. It reports the top 1% of users by migration sentiment. Similarly to the key producers, the results reveal that the top 1% of spreaders retweet 12.6 times more, representing 12.11% of the total retweets. The anti-immigration users represent a majority (70.08%) of the total number of top spreaders. These users generate 21.36% of the total anti-immigration retweets, while key pro-immigration spreaders retweet 6.01% of the total positive retweets on immigration. The top spreaders have no users classified as bots. On the other hand, a larger share of the bottom 99% of the users are bots, namely 0.92%.

**Table 3** - Spreaders grouped by quantile and immigration sentiment in the top 1% of the users.

|  | Users | Retweets | Retweets (%) | Retweets by User | Botness | Influence |
| --- | --- | --- | --- | --- | --- | --- |
| **Top 1%** | 247 | 5,920 | 12.11 | 23.97 | 0.27 | 1.87 |
| **Anti-Immigration** | 176 | 4,149 | 21.36 | 23.57 | 0.31 | 1.88 |
| **Pro-Immigration** | 71 | 1,771 | 6.01 | 24.94 | 0.2 | 1.86 |
| **Bottom 99%** | 24,471 | 42,963 | 87.89 | 1.76 | 0.38 | 1.76 |

*Note:* The top 1% spreaders are selected according to the total number of retweets. Within this top 1%, we identify users classified as 'Anti-immigration' and 'Pro-immigration.' Users' and 'Retweets' columns are counts. 'Retweets by user' is the total count of tweets by category divided by the number of users. 'Botness' is the mean value across the users within that category.

Considering both key spreaders and producers of content in our data, we can identify a sort of 'super users' which are both key spreaders and producers of content. Around 7.28% of the total top 1% producers are also top 1% spreaders. Among these 'super users' (total count = 18), 72.2% are anti-immigration users, further suggesting how the public debate on Twitter around immigration could be disproportionately influenced by a tiny group of people on the platform.

### 4.5 Assessing the speed of the anti (pro) immigration content

Figure 2 displays how fast anti- and pro-immigration tweet cascades reach a certain level measured in median minutes. Figure 2a reports the difference in speed between anti- and pro-immigration retweets across all users. Figure 2b reports this difference excluding users with a botness probability higher than 90%. The results show that anti-immigration content spread, on average, 1.66 times faster than positive pro-immigration tweets. They also revealed that the speed of diffusion differs across the size of tweet cascades. For smaller tweet cascades (<15), there is no difference in speed but for mid-size cascades (15-120), anti-immigration speech spreads consistently faster than pro-immigration tweets. Large cascades (>120) have a similar trend as the mid-range cascades. For some intervals (i.e. around 150), the data is more sparse and less continuous displaying sudden fluctuations. Bots do not seem to exert a major influence on these results. Removing users classified as bots decreases the median speed of anti-immigration content sharing by 12%, but it does not significantly alter the patterns identified above.

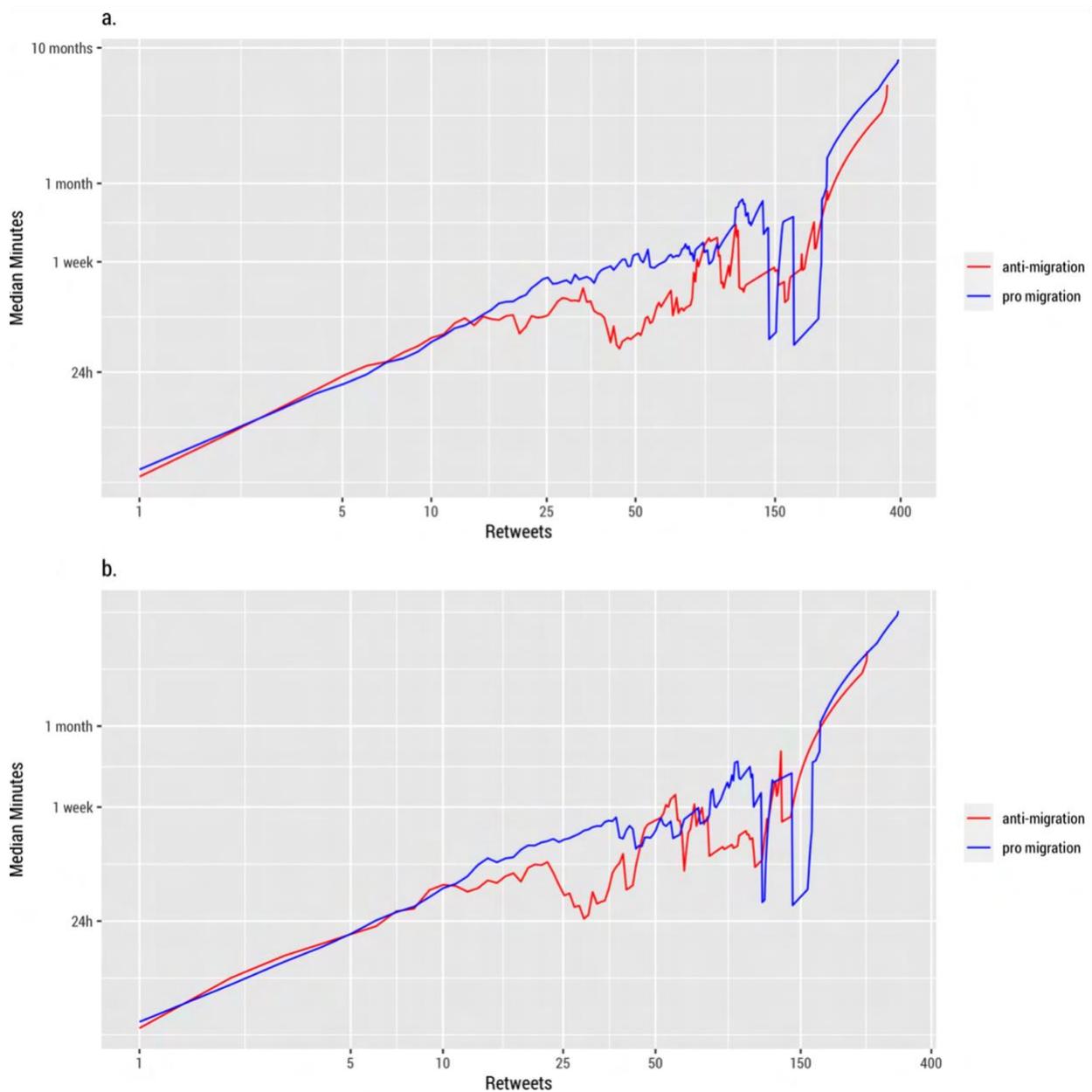

**Figure 2.** Cumulative count of retweets in median minutes on a log-log scale. A) Retweets including all users. B) Retweets without users with a botness probability higher than 90%.

### 4.6 Tweet cascades

Figure 3 displays the CCDFs of pro- and anti-immigration cascades. This informs us how likely is anti- or pro-immigration content to spread by a number of retweets. The higher the value on the y-axis, the higher the probability that an anti- or pro-immigration tweet reaches a certain cascade size i.e. number of retweets. The x-axis of the graph represents the cascade size, which is the number of retweets for a particular tweet related to immigration. The y-axis, on the other hand, represents the complementary cumulative distribution, indicating the probability that a cascade will have at least a certain number of retweets.

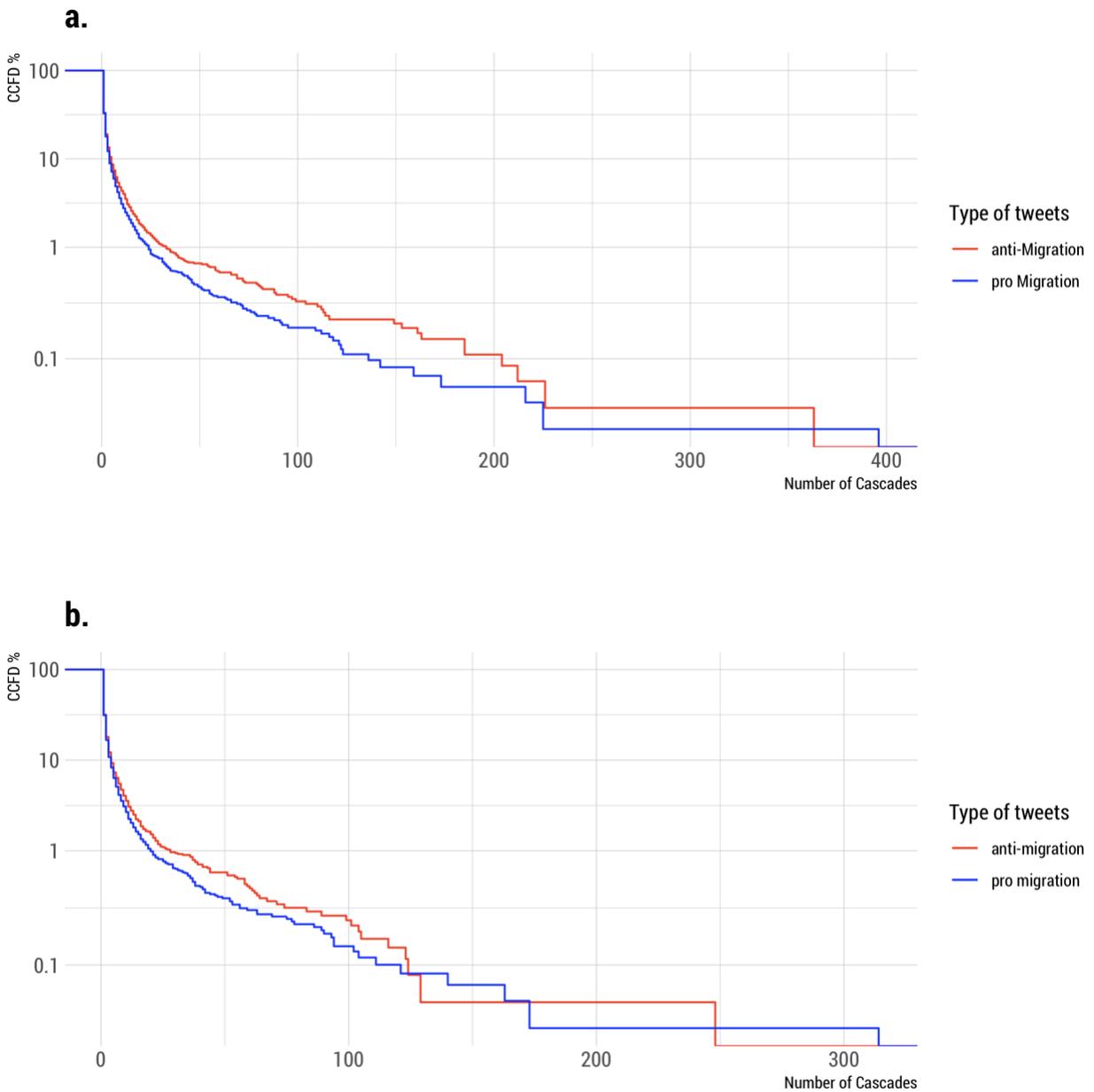

**Figure 3.** Complementary cumulative distribution functions (CCDFs) of pro and anti-immigration cascades. A) Retweets including all users. B) Retweets without users with a botness probability higher than 90%.

In Figure 3, the blue curve represents the CCDF of pro-immigration cascades. Conversely, the red curve portrays the CCDF of anti-immigration cascades. In Figure 3a, we present the two CCDFs based on the complete dataset. In Figure 3b, the same CCDFs are depicted, but with the exclusion of bot users. This comparison reveals interesting patterns in the way anti-immigration tweets circulate among Twitter users. The findings show that, on the whole, anti-immigration tweets tend to have a broader reach across the Twitter user base, especially for cascade sizes smaller than 10. This suggests that anti-immigration content is, on average, more likely to be retweeted, indicating a stronger capacity to spread effectively in the Twittersphere compared to pro-immigration content.

When we consider the inclusion of bots in Figure 3b, the observed difference is somewhat diminished, but their impact does not appear particularly significant. It suggests that while bots do play a role, they don't substantially alter the dynamics of anti-immigration content dissemination.

However, it is important to note that when examining cascade sizes exceeding 120, the observed patterns become less clear and consistent in both figures. This could be due to various factors influencing the behavior and reach of immigration-related content in the online environment.

## 5. Discussion

### 5.1 Key results

Immigration is a significant and divisive topic causing political frictions and social tensions. Social media can enhance these conflicts around immigration. Yet, there is limited quantitative evidence that has used online social networks to investigate the features and dissemination of online discussions about immigration in the UK. Our findings show how the public debate on Twitter around immigration in the UK is largely polarized. The anti-migration network is smaller in size compared to the pro-immigration community, but it is denser suggesting that anti-migration users tend to be more engaged within the community. In the anti-immigration community, users who often retweet are more inclined to engage with fellow frequent retweeters, unlike the pro-immigration network. This finding reinforces the possible positive relationship between user engagement and polarization and could make the anti-migration network more efficient in spreading content internally. A small group of influential users, especially within the anti-immigration network, is responsible for a significant portion of the production and dissemination of polarizing tweets about immigration. Our data shows that only 1% of the producers account for 16% of the total tweets in our dataset. This trend is more pronounced among the anti-immigration community, where slightly over 100 individuals generate 23.18% of the total tweets opposing immigration. Likewise, the top 1% of the spreaders are responsible for 12% of the total retweets, and within the anti-immigration community, the top 1% of spreaders account for 21.36% of the total anti-immigration retweets, far more than the key pro-immigration users. The research reveals that tweets with negative sentiment towards immigration spread 1.66 times faster than positive content. The size of the anti-immigration tweet cascades is systematically larger than the pro-immigration tweets suggesting a higher engagement of anti-immigration messages across time.

**5.2 Implications**

The extent of the polarization in the online public debate on immigration-related issues in the UK could enhance online violence ([Williams et al., 2020](#)) which can ultimately trickle down to physical actions towards migrants and minorities ([Müller & Schwarz, 2021](#)). Yet, a casual-effect relationship between online anti-migration sentiment and racially motivated physical crimes still needs to be established. Our findings show the existence of polarized communities on both pro and anti-immigration stances which can inform existing policymakers to design tools to mitigate polarization on both sides. Indeed, new evidence suggests that online extreme views might feed into each other ([Bail et al., 2018](#); [Törnberg, 2022](#)), yet future studies should explore this thesis in the context of the immigration debate. The content within these polarized communities is largely generated by a small share of the total active users, especially in the anti-immigration community. Our finding is consistent with previous research on politics ([Grinberg et al., 2019](#)) and the COVID-19 anti-vaccine debate ([Nogara et al., 2022](#)) yet it was not proven for the immigration debate. This implies that a small but determined group of people can vastly affect the online public sentiment on immigration. The significance of this discovery lies in its policy implications, as it has the potential to shift content moderation efforts from monitoring a large cohort of users to a smaller but over-influential subset. This is particularly relevant given the ongoing challenges with content moderation policies on major social media platforms ([Gillespie, 2020](#)). Lastly, this novel insight contributes to a broader literature on how users engage in the dissemination of hateful content online.

The study highlights the rapid spread of anti-immigration content compared to pro-immigration content. Our findings underline the urgent need to take swift action to curb online abuses, particularly during specific events that historically trigger the spread of hate speech ([Lupu et al., 2023](#)). Additionally, given the relevance of immigration among the UK public, anti-immigration propaganda has the potential to significantly influence political outcomes. Previous studies have suggested that the dissemination of highly partisan and misleading content may have affected last-minute electoral decisions ([Hopkins & Mutz, 2022](#)), particularly in highly contested areas ([Howard et al., 2017](#)). In the context of health pandemics, the speed at which content spreads can significantly impede public efforts to contain them, emphasizing the importance of mitigating the rapid spread of anti-immigration content ([Borges do Nascimento et al., 2022](#)). The research also expands the existing broader literature on content dissemination within a polarized public debate.

The research findings suggest a connection between polarization, content engagement, and the quick dissemination of immigration-related content on social media.

The anti-immigration community demonstrates higher engagement levels and potential coordination in quickly spreading anti-immigration propaganda, contributing to the polarization among users. These findings align with previous studies indicating that social media platforms, by prioritizing user engagement, can foster polarization and conflicts (Finkel et al., 2020). Policymakers should carefully evaluate how the attention-seeking mechanisms of online social networks impact harmful actions towards migrants. Further research is needed to establish a causal relationship between polarization, content engagement, and the viral nature of immigration-related content. Content moderation policies should prioritize accuracy, particularly in addressing anti-immigration speech driven by misinformation and misperception, especially during health emergencies like the COVID-19 pandemic.

**5.3 Limitations and Challenges**

The research presents a series of challenges and limitations as well. Twitter data represent a small portion of the broader online discussions on immigration given the total userbase on the platform. Furthermore, Twitter's userbase is generally younger and wealthier compared to the general UK population thus it is not representative (Blank & Lutz, 2017). A potential solution might be to include data from other social media platforms but access to actionable users' data has proven to be challenging (Gibney, 2019; Hegelich, 2020). Furthermore, Blank and Lutz (2017) argue that no social media platform is representative of the overall UK population. Using weights to increase the representativeness of the data is a valid option to explore in future studies. Extending the timeframe of the research would benefit the generalization of the findings. Methodologically, the text classifier can be improved especially for labeling anti-immigration content. The current classifier might lead to an underestimate of the anti (pro) immigration tweets since it might falsely label antagonistic tweets towards immigration as positive, neutral, or unclassified. The false positives generated by wrong labeling might bias other results on speed, key producers and key spreaders. Identifying bots can also be challenging and thus it can impact the estimates of the bots' role in speed and key sources of immigration-related content. Rauchfleisch and Kaiser (2020) studied several limitations around the use of machine-learning approaches to discover bots. A major concern is that machine learning methods could be trained on potentially very different datasets than the one on which they are later used. A better approach should be to train the Birdspotter algorithm with a custom-labeled dataset as done by Cresci et al. (2018) and Echeverría et al. (2018).

# 6. Conclusion

We presented evidence of existing polarization in the online public debate on immigration. We identified anti- and pro-immigration networks on Twitter. The pro-immigration network was found to be 1.69 times larger than the anti-immigration network, although the latter is 2.8 times denser. We also identified the primary generators and disseminators of both anti- and pro-immigration content. Our findings indicated that only 1% of producers and disseminators disproportionately generate respectively 16% and 12% of the total content, especially within the anti-immigration community. Less than 1% of the total key producers and spreaders of content were identified as bots. Our findings also revealed that anti-immigration content spreads 1.66 times quicker than pro-immigration content on Twitter with bots playing a marginal role in changing the pace of the content.

The findings painted a concerning picture of the online discourse surrounding immigration. There is a significant level of polarization, as online communities largely engage within their own echo chambers, fostering a sense of isolation and reinforcing existing beliefs. This circumstance has the potential to further solidify existing perspectives and, at its worst, push current users toward more extreme positions on immigration. The higher density of the anti-immigration network indicates that although more individuals have positive views about immigration, those against have stronger connections within their online community. Within these communities, our findings suggested that by identifying and monitoring highly active users, strategic interventions would potentially achieve significant reductions in online hate content as 1% of the key nodes in the network produce 23% of the anti-immigration content in the UK. Our findings also displayed that anti-immigration content spread faster than pro-immigration content emphasizing the need for systematic tools to avoid the widespread dissemination of harmful messages. Failure to address online anti-immigrant sentiment can have serious consequences, including physical harm to those targeted by prejudice. Therefore, there is a critical need to implement tools that can quickly and effectively prevent online anti-immigration speech on a vast scale. We also showed that bots have a marginal role in the online debate on immigration with no significant impact on the production, diffusion and speed of content.

**Acknowledgments:** We are grateful to Michael Mahony for sharing code for data collection and pre-processing. We are also grateful to Carmen Cabrera-Arnau for reviewing the a draft of the research.

**Author Statement:** Andrea Nasuto was responsible for the conceptualization of the project, methodology, software, validation, formal analysis, investigation, data curation, visualization, writing the original draft, review and editing. Francisco Rowe was responsible for the conceptualization of the project, supervision, review and editing and supervision. Michael Mahony collected the original data.